\documentclass[aps,twocolumn,showpacs,preprintnumbers,nofootinbib,prd,superscriptaddress,groupedaddress,10pt]{revtex4-1}

\makeatletter
\def\l@subsubsection#1#2{}
\def\l@subsubsubsection#1#2{}
\makeatother

\usepackage{graphicx,amssymb,amsmath,amsthm,amsfonts,epsfig,epsf}
\usepackage[usenames]{color}
\usepackage{epstopdf}
\definecolor{darkred}{rgb}{0.5,0,0}

\usepackage{array}
\usepackage{afterpage}
\usepackage{bm}
\usepackage{dcolumn}
\usepackage[latin1]{inputenc}
\usepackage{latexsym}
\usepackage{rotating}
\usepackage{longtable}

\usepackage{enumerate}
\usepackage{tensor,multirow}
\usepackage{url}
\usepackage[linktocpage]{hyperref}
\usepackage{float}
\usepackage{graphicx}
\usepackage{amsmath}

\def\be{\begin{equation}}
\def\ee{\end{equation}}
\newcommand{\beq}{\begin{eqnarray}}
\newcommand{\eeq}{\end{eqnarray}}

\def\ba{\begin{align}}
\def\ea{\end{align}}

\newcounter{mnotecount}[section]

\renewcommand{\themnotecount}{\thesection.\arabic{mnotecount}}

\newcommand{\mnote}[1]%{}%
{\protect{\stepcounter{mnotecount}}$^{\mbox{\footnotesize
$%\!\!\!\!\!\!\,
\bullet$\themnotecount}}$ \marginpar{%\color{red}%
\raggedright\tiny\em
$\!\!\!\!\!\!\,\bullet$\themnotecount: #1} }

\begin{document}

\title{Entropy production from quasinormal modes}

\author{Aron Jansen}
\affiliation{Departement de F\'{i}sica Quantica i Astrof\'{i}sica, Institut de Ci\'{e}ncies del Cosmos, Universitat de Barcelona, Mart\'{i} i Franqu\'{e}s 1, E-08028 Barcelona, Spain}
\email{a.p.jansen@icc.ub.edu}
\author{Ben Meiring}
\affiliation{Rudolf Peierls Centre for Theoretical Physics, University of Oxford, Parks Rd, Oxford OX1 3PJ, UK}
\email{ben.meiring@physics.ox.ac.uk}

\begin{abstract} 

Horizons of black branes have an associated entropy current with non-negative divergence. 
We compute this divergence in a late-time transseries expansion for an inhomogeneous system evolving towards a maximally symmetric asymptotically anti-de Sitter black brane. 
The horizon area equilibrates on half the time-scale set by the dominant quasinormal mode and we find a simple analytic expression for this evolution purely in terms of the background and the quasinormal mode frequencies.
This computation does not require a gradient expansion and is thus non-perturbative in momenta.
We generalize this to include scalar and gauge field matter in any number of dimensions.
Restricting to homogeneous evolution we match and prove earlier numerical work showing that the apparent horizon entropy saturates the area theorem, in that its time derivative periodically vanishes. The same is true for spherically symmetric evolution towards the Schwarzschild black hole. 
\end{abstract}

\maketitle

%%%%%%%%%%%%%%%%%%%%%%%%%%%%%%%%%%%%%%%%%%%%%%%%%%%%%%%%%%%%%%%%%%%%%%%%%%

\section{Introduction}
%%%%%%%%%%%%%%%%%%%%%%%%%%%%%%%%%%%%%%%%%%%%%%%%%%%%%%%%%%%%%%%%%%%%%%%%%%
Black holes are some of the most fascinating objects in physics. They have implications for astrophysics, information theory, quantum gravity, and can be used to study holographic theories at strong coupling~\cite{Maldacena:1997re,Ryu:2006bv}. Recently their behaviour out of equilibrium has become experimentally accessible through the detection of gravitational waves from black hole binary collisions~\cite{Abbott:2016nmj}. While this is arguably the most interesting regime of general relativity, it remains difficult to study. Analytic results are scarce and one often has to resort to intensive numerics. Any analytic inroads we can make into this regime are therefore valuable.

In this letter we present analytic results on the out of equilibrium behavior of the horizon, which was recently imaged for the first time~\cite{Akiyama:2019cqa}. 
It has long been known that the horizon area cannot decrease~\cite{hawking1975}, here we show exactly how it increases as a function of its quasinormal modes (QNMs). 
While we focus on black branes in anti-de Sitter spacetime, the results presented here apply as well to spherical evolution of the Schwarzschild black hole, and could potentially be extended beyond spherical symmetry.

A black hole in equilibrium has entropy given by its area~\cite{tHooft:1993dmi,Susskind:1994vu}.
Out of equilibrium this identification is less clear, but analogous to the second law of thermodynamics the black hole area still cannot decrease~\cite{hawking1975}.
In this context one can define various horizons satisfying this law, all of which coincide with the event horizon in equilibrium. In~\cite{Engelhardt:2017aux} it was argued that out of equilibrium it is the apparent horizon that gives the most natural definition of the entropy.

Dynamical horizons have been studied previously through the fluid gravity correspondence~\cite{Bhattacharyya:2008xc}. In these inhomogeneous systems one can define a local entropy current at either the apparent or event horizon with positive semi-definite divergence \cite{Booth:2010kr,Booth:2011qy,Hubeny:2011hd}. Such approaches necessarily neglect the contributions of gapped QNMs, which we include in our analysis.

In~\cite{Jansen:2016zai} the area of spatially homogeneous horizons was studied numerically, and an analytic form was inferred for their late time behavior, 
\begin{equation} \label{eq:delta_entropy}
\delta S(t) \propto e^{2 \omega^{I} \, t} \left( \cos{(2 \omega^{R} \, t)} + B \right)  \, ,
\end{equation}
where $\omega$ is the quasinormal mode (QNM) that gives the dominant damped oscillation, and superscripts indicate real and imaginary parts.

Curiously, the apparent horizon was found to saturate the area theorem, in that the area never decreases but at times is instantaneously constant, periodically reaching $\delta S'(t) = 0$. 
The value of the parameter $B$ for this to happen is
\begin{equation}\label{eq:BAH}
B_\text{AH} = \frac{\left|\omega\right|}{- \omega^{I}} \, .
\end{equation}
In contrast the event horizon did not show a similar saturation and instead always increased.

We build on this previous work by analytically studying the divergence of entropy currents through a transseries expansion.
Transseries expansions have been explored in the context of general relativity explicitly in \cite{Aniceto:2018uik,Casalderrey-Solana:2018uag} and indirectly in \cite{Casalderrey-Solana:2017zyh,Heller:2013fn}.
This is an expansion about a static solution in the parameter $e^{ -i\, \omega \, t}$, where $\omega$ is a stable QNM frequency with negative imaginary part. As the expansion parameter becomes smaller at later times, we also refer to this series as a late time expansion.
The first order in this series contains precisely all the QNMs, and from the second order on their contribution to the backreaction can be seen.
In this paradigm the amplitude of these modes need not be small, instead the expansion is sensible provided $\operatorname{Im}( \omega)$ is an order one negative number.

We obtain the divergence of the entropy current at second order in this late time expansion, 
proving the numerical results mentioned above and further finding that the event horizon entropy will obey Eq. (\ref{eq:delta_entropy}) with,
\begin{equation}\label{eq:BEH}
 B_{EH} =  \frac{\left|\omega\right|}{-\omega^{I}}  \times  \frac{\left|\omega- i \pi T\right|}{-\left( \omega- i \pi T\right)^I} \, ,
 \end{equation}
 where $T$ is the Hawking temperature in equilibrium.

For homogenous perturbations Eq. (\ref{eq:delta_entropy}) gives the QNM contribution to the entropy even for the other maximally symmetric horizons, the sphere and hyperboloid.
The result does not explicitly depend on the asymptotics of the spacetime.
However there can be other contributions, such as the late time tails in asymptotically flat spacetime, that should be included in the transseries.

Our computation will focus however on evolution of black brane solutions in asymptotically anti de Sitter spacetime, and break spatial homogeneity by allowing fluctuations at any finite momentum. We include contributions from all fluctuation channels, namely sound, shear and tensor fluctuations. 
The results we find for the divergence of the entropy current are very simple and fully determined by the QNM frequencies and properties of the equilibrium solution.

We go to second order in the late time expansion, where the first nontrivial contribution occurs.
The third order is expected to vanish because it by itself cannot be non-negative, but we did not prove this.
Therefore our approximation in principle breaks down at times $t \lesssim \left|3 \operatorname{Im}(\omega_{0})\right|^{-1}$, where $\omega_0$ is the dominant QNM, or perhaps one order later.
However it is often the case that even the first order QNM approximation can describe nonlinear general relativity
well beyond its expected regime of validity, even with large perturbations~\cite{Heller:2013oxa,Heller:2013fn,Heller:2012km}.
We stress that while the numerical work in \cite{Jansen:2016zai} only identified the contribution of the dominant QNM, 
the analytic computation we provide holds for QNMs of any order, they will all contribute in the same way.

We organise this letter as follows.
In section (\ref{sec:Setup}) we set up the problem and define the metric, entropy current and transseries expansion. 
In section (\ref{sec:Results}) we show our main result Eq. (\ref{eq:div_S}) for the divergence of the entropy current due to QNM perturbations in the case of Einstein gravity with negative cosmological constant. 
We explain how the entropy density can be obtained from this expression in the case of homogeneous perturbations. 
In section (\ref{sec:Generalizations}) we generalize our result to Einstein-Maxwell-Scalar theory with arbitrary potential and gauge coupling. 
We give the divergences in full detail in Appendix \ref{sec:matter}.
Finally in section (\ref{sec:discussion}) we discuss the implications and extensions of our analysis.

%%%%%%%%%%%%%%%%%%%%%%%%%%%%%%%%%%%%%%%%%%%%%%%%%%%%%%%%%%%%%%%%%%%%%%%%%%
\section{Setup} \label{sec:Setup}
%%%%%%%%%%%%%%%%%%%%%%%%%%%%%%%%%%%%%%%%%%%%%%%%%%%%%%%%%%%%%%%%%%%%%%%%%%
For simplicity we start with a black brane in anti-de Sitter in $(4+1)$-dimensions. Later we will generalize to arbitrary dimension and also include matter.

We take the ansatz for the metric,
\begin{equation}\label{eq:metricansatz}\begin{split}
ds^2 &= - f dt^2 + 2 dt \left(dr + F^{(x)} dx + F^{(y)} dy \right) + \Sigma^2  g_{i j} \, , \\
g_{i j } &= 
\begin{pmatrix}
e^{-2 B}\cosh{G} & e^{-\frac{1}{2}(B-H)} \sinh{G} & 0 \\
e^{-\frac{1}{2}(B-H)}\sinh{G} & e^{B+H} \cosh{G}& 0 \\
0 & 0 & e^{B-H}
\end{pmatrix} \, ,
\end{split}\end{equation}
where all functions depend on $(t,r,x)$ but not on the remaining coordinates $(y,z)$.
To simplify the resulting equations of motion we choose a particular ansatz such that the determinant of the metric is given by a single function $\sqrt{-g} = \Sigma^3$. 

We further restrict to static, homogeneous solutions at equilibrium so that at infinite $t$ the only nonzero functions will be $f$ and $\Sigma$ which will depend only on $r$.
The equilibrium solution is then simply the AdS-Schwarzschild black brane,
\begin{equation}\label{eq:schw}\begin{split}
f_0 &= r^2\left(1 - \left(\frac{r_h}{r}\right)^4\right) \, , \quad \Sigma_0(r) = r \, , \\
F_{0}^{(x)} &= F_{0}^{(y)} = B_0 = G_0 = H_0 = 0 \, ,
\end{split}\end{equation}
with equilibrium temperature given by $T = f_0^\prime(r_h)/(4\pi)$. This temperature can be expressed purely in terms of $r_h$ but to facilitate an easy generalization we do not use Eq. (\ref{eq:schw}) explicitly, using instead the definition of $T$ along with the fact that $f_0(r_h) = 0$ at the horizon. 
Away from equilibrium this ansatz contains a tensor fluctuation through $H$, a vector or shear fluctuation through $F^{(y)}$ and a scalar or sound fluctuation through $F^{(x)}$.

%%%%%%%%%%%%%%%%%%%%%%%%%%%%%%%%%%%%%%%%%
\subsection{Entropy currents}
%%%%%%%%%%%%%%%%%%%%%%%%%%%%%%%%%%%%%%%%%

Out of equilibrium the identification of entropy with a horizon surface becomes ambiguous as multiple distinct codimension-2 surfaces exist on which to define one. 
We will consider two such surfaces, the event horizon (EH) and the apparent horizon (AH).
The event horizon is a null surface defined by its normal vector $n^M$, where $M$ runs over all the coordinates,
while the apparent horizon is a spacelike surface on which the geodesic expansion $\theta$ vanishes
\footnote{There is a subtlety here in that this definition depends on how the full spacetime is foliated by spacelike surfaces, 
however here the late time expansion clearly singles out constant time slices as the natural foliation.},
\begin{equation}\label{eq:horizonconditions}
\text{AH}: \theta |_{r_\text{AH}} = 0 \, , \quad \text{EH}: n_M n^M  |_{r_\text{EH}} = 0 \, .
\end{equation}
In the infinite $t$ limit these surfaces align and are given radially by the zero of the blackening function $f$. 

Our ansatz Eq. (\ref{eq:metricansatz}) preserves the residual gauge freedom associated with a choice of radial shift $r \rightarrow r + \xi(t,x)$. This is routinely used in numerical evolution schemes to fix the position of a locally determined apparent horizon, while doing the same for a globally determined event horizon is usually not possible. 
In our setup however this is no issue as we assume knowledge of our final state and expand our solution backwards in time in a controlled way.
 We will fix this gauge implicitly by imposing either condition of Eq. (\ref{eq:horizonconditions}) at a fixed constant radius radius $r_\text{AH}$ or $r_\text{EH}$ which is independent of the other coordinates.

For each horizon we can define an entropy current, following \cite{Booth:2010kr}, through its normal vector $v_M$, which through our chosen radial gauge is simply $v = \partial_r$. The entropy current is then given by,
\begin{equation}\label{eq:entropycurrent}
S^\mu = s \frac{v^\mu}{v^t} \, ,
\end{equation}
where $s = \sqrt{-g}/4$ is the entropy density evaluated at the appropriate horizon, and $\mu$ runs over all but the radial coordinate~\cite{Bhattacharyya:2008xc,Herzog:2016hob}. 
The divergence of this entropy current is constrained to be non-negative through the area theorems \cite{Booth:2010kr}, 
\begin{equation} 
\partial_\mu S^\mu \geq 0 \, ,
\end{equation}
which is true for both the event and apparent horizon.

To map this current from the horizon to the boundary we simply map the radial coordinate, leaving the others fixed, as was done in~\cite{Booth:2010kr,Bhattacharyya:2008xc}.

%%%%%%%%%%%%%%%%%%%%%%%%%%%%%%%%%%%%%%%%%
\subsection{Late time expansion}
%%%%%%%%%%%%%%%%%%%%%%%%%%%%%%%%%%%%%%%%%
Following \cite{Aniceto:2018uik}, we write the late time expansion as a transseries in $t$, or a power series in
\begin{equation}
e_n(t,x) = e^{- i \left( \omega_n t - k_n x \right)}  \, .
\end{equation}

In this work we only go to second order and use a simplified notation.
We expand any function in the metric as follows,
\begin{equation}\label{eq:latetimenew}\begin{split}
&g(r,t,x) = g(r) + \sum_{n=0}^\infty  \left\{ g_n(r) e_n(t,x)+ g_{\bar{n}}(r) e_{\bar{n}}(t,x) \right\} \\
&+ \Bigg[ \sum_{n=0}^\infty  \left\{ g_{n n} e_n^2 + \left(g_{n \bar{n}} + g_{\bar{n} n}\right) e_n e_{\bar{n}}   +g_{\bar{n} \bar{n}} e_{\bar{n}}^2 \right\} \\
&+\sum_{n \neq m=0}^\infty \sum_{a \in \{n, \bar{n} \} } \sum_{b \in \{m,\bar{m}\}} g_{a b} e_a e_b\Bigg]+ \mathcal{O}(e^3)\, , 
\end{split}\end{equation}
where the coefficients $g$ depend only on $r$, we suppress the $(t,x)$ dependence of $e$, and define $k_{\bar{n}} \equiv - k_n$,  $\omega_{\bar{n}} \equiv -  \bar{\omega}_n $, where $\bar{\omega}_n$ is the complex conjugate of ${\omega}_n$.
\footnote{Both $\omega_{n}$ and $-\bar{\omega}_{n}$ are QNMs. To avoid double counting in Eq. (\ref{eq:latetimenew}) we take the indices $n$ and $m$ to sum over only QNMs with $\omega^{R}>0$.  We could have equally chosen to sum over only the $\omega^{R}<0$ modes.} 
With this convention of using the same symbol $g$ without indices for the equilibrium value, the number of indices is equal to the order in the expansion. Higher order contributions can be included in a similar fashion to \cite{Aniceto:2018uik}.

The first term $g(r)$ is simply the infinite time equilibrium solution, 
note that in what follows we will suppress the subscript $0$ as given in Eq. (\ref{eq:schw}).
The first order correction gives linearized perturbations about equilibrium, contributions which are identical to quasinormal modes. 
To preserve the reality of the solution we must include each mode in complex conjugate pairs.

These QNMs come in three decoupled channels, tensor, shear and sound with helicities $2$, $1$ and $0$ respectively. 
It is convenient to work with gauge-invariant combinations of these perturbations given respectively by,
\begin{equation} \label{eq:GImodes}\begin{split}
X_{n} &\equiv H_n \, , \\
Y_{n} &\equiv k_n F^{(y)}_n + \omega_n \Sigma^2 G_n \, , \\
Z_{n} &\equiv 4 k_n \omega_n F^{(x)}_n + 2 k_n^2 \left( \frac{f^\prime}{\Sigma^\prime} \Sigma_n- f_n \right) \\
&+ \Sigma \left(k_n^2 \frac{ f^\prime}{\Sigma^\prime} - 6 \omega_n^2 \Sigma \right) B_n\, .
\end{split}\end{equation}
The modes in Eq. (\ref{eq:GImodes}) each satisfy their own decoupled QNM equation.
Under simple coordinate redefinitions these equations are found to be the same as those studied in \cite{Kovtun:2005ev}. 
The usual physical boundary conditions, where the fluctuations die off at infinity and are ingoing at the horizon, allow for a countably infinite set of frequencies $\omega_n$, which are all included in this first order.

As linear perturbations around equilibrium, these only depend on the equilibrium solution where the horizons coincide, $r_{AH} = r_{EH} = r_h$.
The amplitude and phase of a given mode at the horizon remain undetermined by the QNM equations and can be considered initial data. 
We parameterise these constants as,
\begin{equation} \label{eq:horizonValues}
X_n(r_h) = \mathcal{X}_n e^{i \alpha_n^\mathcal{X}/2} \, ,
\end{equation}
where by the reality of the solution $X_{\bar{n}}= \bar{X}_{n}$ with analogous definitions holding for $Y$ and $Z$.

Finally and most crucially we include second order contributions in $e_n$. 
These terms can be thought of as the interactions of pairs of QNMs, and give the leading backreaction due to the perturbations. 
On the third line of Eq. (\ref{eq:latetimenew}) we see the contribution of a single mode, which interacts with its complex conjugate. 
If we had artificially excited only a single mode we could collapse the sum over $n$ and stop here. 
Generically however an ensemble of modes are excited and this sum will include interactions between every different pair, given on the last line in Eq. (\ref{eq:latetimenew}). 
In this analysis we neglect interactions of 3 and higher modes, which can in principle be included in a similar way. 

An important point in the calculation that follows is that the exponentials $e_n$ are linearly independent. 
From this property we will be able to separate the equations of motion out into independent ODEs in $r$ for the coefficients $g_{n m}(r)$, 
sourced by the first order contributions.
These equations further reduce to algebraic relations when evaluated on either the event- or apparent horizon, which allows us to write the divergence of the entropy currents purely in terms of the horizon values defined in Eq. (\ref{eq:horizonValues}).

%%%%%%%%%%%%%%%%%%%%%%%%%%%%%%%%%%%%%%%%%%%%%%%%%%%%%%%%%%%%%%%%%%%%%%%%%%
\section{Results} \label{sec:Results}
%%%%%%%%%%%%%%%%%%%%%%%%%%%%%%%%%%%%%%%%%%%%%%%%%%%%%%%%%%%%%%%%%%%%%%%%%%
We now want to compute the divergence of the entropy current. Before going into the concrete calculation, we can argue very generally that the first order contribution must vanish. 
First order contributions from Eq. (\ref{eq:latetimenew}) take the form of damped oscillations. 
These oscillating functions can not be non-negative, so that any non-negative quantity built from them must vanish to preserve non-negativity.
As shown below, the interaction of modes can cancel out this oscillation at second order.

Following the procedure outlined in section (\ref{sec:Setup}), we can write the second order of the divergence of the entropy current into the following form, 
\begin{widetext}
\begin{equation}\label{eq:div_S}
\partial_\mu S^\mu = \sum_{\mathcal{H} \in \left\{\mathcal{X}, \mathcal{Y}, \mathcal{Z} \right\}}  \sum_{n, m=0}^\infty  A_{n  m} e^{\left(\omega_n + \omega_m \right)^I t} \left[ \cos \left(\Omega^+_{n m} + \delta^+_{n m} \right) + C_{n m}\cos \left(\Omega^-_{n m} + \delta^-_{n m} \right) \right]  + \mathcal{O}(e^3)\, ,
\end{equation}
\end{widetext}
with 
\begin{equation}
\Omega^\pm_{n m} = \left(k_n \pm k_m \right) x- \left(\omega_n \pm \omega_m \right)^R t \, .
\end{equation}
The first sum runs over the different fluctuation channels,
then within each channel indicies $m$ and $n$ sum over QNMs with $\omega^{R}>0$.
While we have simultaneously included all the channels we observe that each channel in Eq. (\ref{eq:div_S}) decouples completely with no interactions between different channels at this order. The parameters $A_{n m}$ and $\delta^\pm_{n m}$ and the frequencies $\omega_n$ depend on the channel, but for notational simplicity we suppress the indices of $\mathcal{X}$, $\mathcal{Y}$ and $\mathcal{Z}$.

Each term in this sum has an amplitude $A_{n m}$ and phases $\delta_{n m }^\pm$ that depend on the initial condition through the corresponding QNMs.
The more interesting parameter is $C_{n m}$, which is related to what was called the damping shift in~\cite{Jansen:2016zai}.
In the simplest case of $n = m$ the cosine that multiplies it simplifies to one, and this parameter gives a constant term that suppresses the other oscillatory term. 
This suppression of the oscillation is crucial to having a positive definite quantity, in particular in the diagonal terms it has to be greater or equal to one. 
To distinguish it from the slightly differently defined damping shift in homogeneous scenarios~\cite{Jansen:2016zai} we will refer to the parameter $C_{n m}$ as the \textit{oscillation suppression}.

All parameters in Eq. (\ref{eq:div_S}) are determined from the equations of motion.
By imposing either horizon condition Eq. (\ref{eq:horizonconditions}) order by order, the Einstein equations $E_{\mu\nu} = 0$ give relations between the various metric functions at the horizon that allow us to express Eq. (\ref{eq:div_S}) fully in terms of the QNM frequencies and the horizon values Eq. (\ref{eq:horizonValues}).
In particular the only necessary second order constraint comes from solving the unexpanded $E_{rr}$ for $\partial_r^2 \Sigma$, inserting that into $E^r_t$ and expanding the result to second order.

To present the results compactly we define,
\begin{equation}\label{eq:rescalings}\begin{split}
k_n &= \sqrt{(d-1) \pi T r_h/2} \,  q_n \, , \quad \omega_n =  \pi T \lambda_n \, , \\
q_\pm &= 1/2 (q_n \pm q_m) \, , \quad \lambda_\pm = 1/2 (\lambda_n \pm \lambda_m) \, ,
\end{split}\end{equation}
where $d=4$ is the number of dimensions in the boundary, which we keep for later generalization.
For notational simplicity we do not explicitly write the $n, m$ dependence of $\lambda_\pm$ and $q_\pm$.

The oscillation suppression $C_{n m}$ is universal across the different channels of fluctuations, 
\begin{equation} \label{eq:oscillationSuppression}
C_{n m} = \begin{cases}\displaystyle
\frac{1+q_+^2}{1+q_-^2}  \quad &\text{(AH)} \, , \\
\displaystyle
\sqrt{\frac{\left(\lambda_+^{R} \right)^2+\left(\lambda_+^{I} -  1 \right)^2}{  \left(\lambda_-^{R}\right)^2+\left(\lambda_+^{I} -  1 \right)^2} } \quad &\text{(EH)} \, .
\end{cases}
\end{equation}
Note that $C_{n n} \geq 1$, since then $q_- = \lambda_- = 0$, which directly implies positivity of the diagonal contributions to the divergence of the entropy current.
Including the mixed terms checking positivity is more involved, see Appendix \ref{sec:Positivity} for a proof of positivity of the divergence of the entropy current due to the lowest two tensor fluctuations.
For the other cases we have done extensive numerical checks showing that they are indeed always positive. 

At zero momentum $C_{n n} = 1$ for the apparent horizon, which then saturates the area theorem. 
Interestingly $C_{n n}$ for the event horizon is momentum independent while for the apparent horizon it is frequency independent.

We recover the previous numerical results of Eqs. (\ref{eq:BAH}) and (\ref{eq:BEH}) by setting $m=n$, setting the momenta to zero and integrating Eq. (\ref{eq:div_S}) to go from $\partial S$ to $S$ itself (this gives an extra factor $\left|\omega\right| / (- \omega_I)$).

The amplitude and phase of each mode will depend on initial data, but for a given mode the ratio of the amplitudes and the difference of the phases associated with the apparent and event horizon will be independent of these choices. 
For the amplitudes we have the general form,
\begin{equation}\label{eq:amplitudes}
A_{n m} = \frac{r_h^{d-5}}{8 \pi T} c_n c_m \begin{cases}\displaystyle
\frac{1}{ (1 + q_+^2)}  \quad &\text{(AH)} \, , \\
\displaystyle
\frac{1}{ \left| \lambda_+ - i \right|}  \quad &\text{(EH)} \\
\end{cases}
\end{equation}
where the $c_n$ depend on the channel as,
\begin{equation}\label{eq:amplitudescs}
c_n = \begin{cases}\displaystyle
\pi T r_h^2 \left|  \lambda_n \right| \mathcal{X}_n \quad &\text{tensor} \, ,\\
\mathcal{Y}_n \quad &\text{shear} \, ,\\
\sqrt{\frac{d-2}{8(d-1)}}\frac{\left|  \lambda_n \right| \mathcal{Z}_n}{ \pi T \left|  \lambda_n^2 -  q_n^2 \right|} \quad &\text{sound} \, .
\end{cases}
\end{equation}
We stress again that there is no mixing between the channels.

The phases $\delta_{n m}^{\pm}$ are somewhat complicated and given in Appendix \ref{sec:matter}, but the difference between the phases evaluated at the event and apparent horizons for all channels is simple and given by,
\begin{equation}\label{eq:phases}\begin{split}
\delta_{n m}^{\pm \text{ EH}} - \delta_{n m}^{\pm \text{ AH}}  = & \arctan \left(\frac{\lambda_{\pm}^{R}}{\lambda_{+}^{I}-1} \right) \, .
\end{split}\end{equation}

%%%%%%%%%%%%%%%%%%%%%%%%%%%%%%%%%%%%%%%%%%%%%%%%%%%%%%%%%%%%%%%%%%%%%%%%%%
\section{Generalizations} \label{sec:Generalizations}
%%%%%%%%%%%%%%%%%%%%%%%%%%%%%%%%%%%%%%%%%%%%%%%%%%%%%%%%%%%%%%%%%%%%%%%%%%
While in the previous section we have worked in specifically $d=4$, the results are given above for arbitrary dimension.\footnote{We have done the computations for $d = 4, ..., 10$ and extrapolated the results.}
The only dependence occurs in the rescaled momentum,  the overall power of $r_h$ in Eq. (\ref{eq:amplitudes}) (which can be obtained from dimensional analysis), and a numerical prefactor in the $c_i$ for sound.

We further generalize our analysis to a broad class of Einstein-Maxwell-scalar theories with action\footnote{For a derivation of master equations in exactly this setting, which are ideal to compute QNMs, see~\cite{Jansen:2019wag}.},
\begin{equation}\label{eq:EMSaction}
S = \!\int\!\! d^{d+1} x \sqrt{-g} \!\left(\!R\! - 2\! \Lambda\! - \frac{1}{2} (\partial \phi)^2 \!-\! \frac{Z(\phi)}{4}  F^2 \!-\! V(\phi) \right) \, ,
\end{equation}
with the same metric as defined in Eq. (\ref{eq:metricansatz}) and where the gauge field $F = d a$ has non-zero $(t,r,x,y)$ components.

The matter fields give rise to one more gauge invariant in the shear channel and two more in the sound channel given respectively by,
\begin{equation}\label{eq:genGI}\begin{split} 
U_n &\equiv a^{(y)}_{n} \, , \\
V_n &\equiv 2( k a^{(t)}_{n} + \omega a^{(x)}_{n}) -k \frac{a^{(t)}{}^\prime}{\Sigma^\prime}  \left(\Sigma B_{n} +2\Sigma_{n}\right) \, , \\
W_n &\equiv 2 \phi_{n}  - \frac{\phi^\prime}{\Sigma^\prime}  \left(\Sigma B_{n} +2\Sigma_{n}\right) \, ,
\end{split}\end{equation}
the matter gauge invariants of Eq. (\ref{eq:GImodes}) remain unchanged. We define the horizon values of $U_n$, $V_n$ and $W_n$ in a similar way to Eq. (\ref{eq:horizonValues}).
Following the same steps as outlined in sections (\ref{sec:Setup}) and (\ref{sec:Results}) the same calculation can be performed. The different channels still decouple in Eq. (\ref{eq:div_S}), however we will now have multiple gauge invariants in each channel and cross terms between these gauge invariant modes do exist.

For clarity the new form of the divergence of the entropy current, including explicit forms for the amplitudes and apparent horizon phases is given Appendix \ref{sec:matter}. 
Here we give a more qualitative overview.

Instead of Eq. (\ref{eq:rescalings}) we now rescale the momentum as
$k_n = \sqrt{(d-1) \pi T \Sigma_h \Sigma_h' / 2} \, q_n $
and replace the $r_h$ in Eqs. (\ref{eq:amplitudes}) and (\ref{eq:amplitudescs}) by $\Sigma_h$, where the subscript $h$ indicates evaluation at $r_h$.

After these trivial changes the oscillation suppression as given in Eq. (\ref{eq:oscillationSuppression}) remains identical for all channels with the exception of the $\mathcal{Z}_n \mathcal{Z}_m$ contribution.
Furthermore both the ratio of amplitudes and difference in phases between the event and apparent horizon, as given in Eqs. (\ref{eq:amplitudes}) and (\ref{eq:phases}), remain the same without exception.

Outside of the replacements mentioned above, the tensor channel contribution $\mathcal{X}_n \mathcal{X}_m$ remains identical.

The shear channel contribution from $\mathcal{Y}_n \mathcal{Y}_m$ remains similarly unchanged, 
however we find an additional and near identical contribution from the gauge invariant $U_n$ proportional to $ \mathcal{U}_n \mathcal{U}_m$. 
There is no cross term in $\partial_{\mu} S^{\mu}$ proportional to $\mathcal{Y}_n \mathcal{U}_m$.

The sound channel is more involved. In addition to two new quadratic contributions from gauge invariants $V_n$ and $W_n$ we find additional cross terms proportional to $\mathcal{V}_{n} \mathcal{Z}_{m}$ and $\mathcal{W}_{n} \mathcal{Z}_{m}$. 
There is no cross term due to interactions between the $V_n$ and $W_n$ modes.

The oscillation suppression $C_{n m}$ of the $\mathcal{Z}_n \mathcal{Z}_m$ contribution is enhanced by a multiplicative correction ${C_{n m}^c \geq 1}$ due to the time component of the gauge field.
The amplitude of this contribution also changes, see Appendix \ref{sec:matter} for this and all other details. We have verified numerically that $\partial_\mu S^\mu \geq 0$ for both horizons in all channels.

We finally remark that while the QNM equations all decouple at zero momentum, the cross term between the scalar field and the metric, proportional to $\mathcal{W}_{n} \mathcal{Z}_{m}$, does not vanish in this limit, as can be seen in the Appendix. 

%%%%%%%%%%%%%%%%%%%%%%%%%%%%%%%%%%%%%%%%%%%%%%%%%%%%%%%%%%%%%%%%%%%%%%%%%%
\section{Discussion} \label{sec:discussion}
%%%%%%%%%%%%%%%%%%%%%%%%%%%%%%%%%%%%%%%%%%%%%%%%%%%%%%%%%%%%%%%%%%%%%%%%%%
We have obtained explicit expressions for the divergence of the entropy currents associated with the event and apparent horizon to second order in a late time transseries expansion, allowing for the interactions of two different modes. 

At first order no contribution is allowed due to the non-negativity of entropy growth, meaning that the entropy will equilibrate at half the time scale set by the dominant QNM. 

At third order we expect no contributions by the same argument, although it is not strictly necessary for the third order contribution to be non-negative by itself.
For tensor perturbations we have checked explicitly that to $\partial_{\mu} S^{\mu}$ has no $\mathcal{O}(e_{n}^{3})$ correction at third order.

We find in this case that the second order tensor fluctuation satisfies the same equation as the first order tensor QNM equation, but with twice the frequency. Since this is not generically a QNM frequency itself, there will be no nontrivial solution to this equation with physical boundary conditions. This second order fluctuation then gives a vanishing contribution to the divergence of the entropy current, setting the third order contribution to zero. Optimistically our expansion would only receive non-trivial corrections at $\mathcal{O}(e_{n}^4)$ more generally but this warrants an explicit check. 

In contrast to the analysis performed in~\cite{Bhattacharyya:2008xc,Booth:2011qy} our use of the transseries expansion allowed use to avoid explicitly truncating our expansion in gradients. Therefore our result is accurate to all orders of momenta provided one is able to compute $\omega(k)$. Furthermore our expressions can be applied equally to hydrodynamic and gapped QNMs.

In our setup we have only allowed for spatial dependence along the $x$ coordinate. While one can do this at linear level without loss of generality, this is not the case at second order and it remains to be seen what the contribution of modes with momenta along the other directions would be. 

We have focused on asymptotically anti-de Sitter black branes but similar results can be obtained for spherical or hyperbolic black objects, or for de Sitter or flat asymptotics. In fact the homogeneous, zero momentum limit of the results presented here apply directly to those cases, even in the presence of matter fields. 
In asymptotically flat spacetimes there is the complication of late time tails, or polynomial decays which are present on top of the QNMs. 
These we have not treated, but our result still applies to the entropy generated by the interaction of QNMs separately. A further analysis could in principle include these effects for a more reliable description of the evolution of the horizon.

%%%%%%%%%%%%%%%%%%%%%%%%%%%%%%%%%%%
 \begin{acknowledgments}
 %%%%%%%%%%%%%%%%%%%%%%%%%%%%%%%%%%%
We thank Jorge Casalderrey Solana, Roberto Emparan, Michal Heller, Chris Herzog, Javier Magan, Wilke van der Schee, Michal Spalinski and Andrei Starinets for comments on an earlier version of this work.
The work of AJ is supported by ERC Advanced Grant GravBHs-692951. BM is supported by the Skye Foundation and the Oppenheimer Memorial Trust. 
 \end{acknowledgments}
%%%%%%%%%%%%%%%%%%%%%%%%%%%%%%%%%%%%%%%%%%%%%%%%%%%%%%%%%%%%%%%%%%%%%%%%%%%%%

%%%%%%%%%%%%%%%%%%%%%%%%%%%
\bibliography{entropyEquilibrationv2}
%%%%%%%%%%%%%%%%%%%%%%%%%%%

%%%%%%%%%%%%%%%%%%%%%%%%%%%%%%%%%%%%%%%%%%%%%%%%%%%%%%
\begin{appendix}
%%%%%%%%%%%%%%%%%%%%%%%%%%%%%%%%%%%%%%%%%%%%%%%%%%%%%%

\clearpage
%%%%%%%%%%%%%%%%%%%%%%%%%%%
\section{Solution in full detail} \label{sec:matter}
%%%%%%%%%%%%%%%%%%%%%%%%%%%

In this section we give our most general result in a self contained and fully explicit form for a theory given by Eq. (\ref{eq:EMSaction}). The complete expression for the divergence of the entropy current is given by,
\begin{equation}\label{eq:divSmatter}\begin{split}
\partial_\mu S^\mu &= \sum_{n, m} \mathcal{A}_{\mathcal{X}_n \mathcal{X}^\prime_m} + \sum_{\mathcal{H}, \mathcal{H}^\prime \in \left\{ \mathcal{U}, \mathcal{Y} \right\}} \sum_{n, m} \mathcal{A}_{\mathcal{H}_n \mathcal{H}^\prime_m} \\
&+ \sum_{\mathcal{H}, \mathcal{H}^\prime \in \left\{ \mathcal{V}, \mathcal{W}, \mathcal{Z} \right\}} \sum_{n, m} \mathcal{A}_{\mathcal{H}_n \mathcal{H}^\prime_m} \, ,
\end{split}\end{equation}
where we have expanded the sum over channels and defined
\begin{equation}\begin{split}
\mathcal{A}_{\mathcal{H}_n \mathcal{H}^\prime_m} &= e^{\left(\omega_{\mathcal{H}_n} + \omega_{\mathcal{H}^\prime_m} \right)^I t} A_{\mathcal{H}_n \mathcal{H}^\prime_m}  \Big[ \cos \left(\Omega^+_{\mathcal{H}_n \mathcal{H}^\prime_m} + \delta^+_{\mathcal{H}_n \mathcal{H}^\prime_m} \right) \\
&+ C_{\mathcal{H}_n \mathcal{H}^\prime_m} \cos \left(\Omega^-_{\mathcal{H}_n \mathcal{H}^\prime_m} + \delta^-_{\mathcal{H}_n \mathcal{H}^\prime_m} \right) \Big]\, ,
\end{split}\end{equation}
and
\begin{equation}
\Omega^\pm_{\mathcal{H}_n \mathcal{H}^\prime_m} = \left(k_{\mathcal{H}_n} \pm k_{\mathcal{H}^\prime_m} \right) x- \left(\omega_{\mathcal{H}_n} \pm \omega_{\mathcal{H}^\prime_m} \right)^R t \, .
\end{equation}

Two of the cross terms vanish:
\begin{equation}
A_{\mathcal{Y}_n \mathcal{U}^\prime_m} = A_{\mathcal{V}_n \mathcal{W}^\prime_m} = 0 \, ,
\end{equation}
the other contributions are all nonzero.

We have computed this expression for the event- and apparent horizon, which we denote by EH and AH respectively.
The difference between the two is identical for each contribution, and summarized as 
\begin{equation}\begin{split}
C_{\mathcal{H}_n \mathcal{H}^\prime_m} &= C^c_{\mathcal{H}_n \mathcal{H}^\prime_m} \times \begin{cases}\displaystyle
\frac{1+q_+^2}{1+q_-^2}  \quad &\text{(AH)}  \\
\displaystyle
\sqrt{\frac{\left(\lambda_+^{R} \right)^2+\left(\lambda_+^{I} -  1 \right)^2}{  \left(\lambda_-^{R}\right)^2+\left(\lambda_+^{I} -  1 \right)^2} } \quad &\text{(EH)} 
\end{cases} \\
A_{\mathcal{H}_n \mathcal{H}^\prime_m} &= \frac{\Sigma_h^{d-5}}{8 \pi T} c_{\mathcal{H}_n \mathcal{H}^\prime_m} \begin{cases}\displaystyle
\frac{1}{ (1 + q_+^2)}  \quad &\text{(AH)}  \\
\displaystyle
\frac{1}{ \left| \lambda_+ - i \right|}  \quad &\text{(EH)} \\
\end{cases} \\
\delta^{\pm,\text{EH}}_{\mathcal{H}_n \mathcal{H}^\prime_m} &- \delta^{\pm,\text{AH}}_{\mathcal{H}_n \mathcal{H}^\prime_m} =
 \arctan \left(\frac{\lambda_{\pm}^{R}}{\lambda_{+}^{I}-1} \right) \, .
\end{split}\end{equation}

Here we have defined 
\begin{equation}\begin{split}
k_n &= \sqrt{(d-1) \pi T \Sigma_h \Sigma_h^\prime /2} \,  q_n \, , \quad \omega_n =  \pi T \lambda_n \, , \\
q_\pm &= 1/2 (q_n \pm q_m) \, , \quad \lambda_\pm = 1/2 (\lambda_n \pm \lambda_m) \, ,
\end{split}\end{equation}
and we have omitted the dependence of the frequency and momenta on the corresponding gauge invariant purely for notational simplicity.

The oscillation suppressions are almost fully universal, with
\begin{equation}\begin{split}
C^c_{\mathcal{H}_n \mathcal{H}^\prime_m} &= 1 \, , \quad \mathcal{H},\mathcal{H}^\prime \neq \mathcal{Z},\mathcal{Z} \, ,\\
C^c_{\mathcal{Z}_n \mathcal{Z}^\prime_m} &= \sqrt{\frac{\mathcal{R}^2 \left| \lambda_n \right|^2 \left| \lambda_m \right|^2+q_n q_m \mathcal{Q}\left(q_n q_m \mathcal{Q} + 2 \mathcal{R} \nu_{n m}^+ \right)}{\mathcal{R}^2 \left| \lambda_n \right|^2 \left| \lambda_m \right|^2+q_n q_m \mathcal{Q}\left(q_n q_m \mathcal{Q} + 2 \mathcal{R} \nu_{n m}^- \right)}} \, ,
\end{split}\end{equation}
where $\mathcal{Q}$ and $\mathcal{R}$ depend on the background as
\begin{equation}\begin{split}
\mathcal{Q} &= 32 (d-1)  \pi T \Sigma_h \Sigma_h^\prime  Z_h \left(a_h^{(t)}{}^\prime\right)^2\ \, ,  \\
\mathcal{R} &= \Sigma_h^2 \left(  -2 V_h' + Z_h' \left(a_h^{(t)}{}^\prime\right)^2 \right) \\
&+ 128 (d-1) (d-2) \left(\pi T \Sigma_h^\prime\right)^2 \, , \\
\end{split}\end{equation}
and we further define several combinations of frequencies $\nu_i$ and of frequencies and momenta $\kappa_i$ as
\begin{equation}\begin{split}
\nu_{n m}^\pm &= \lambda_n^R \lambda_m^R \pm \lambda_n^I \lambda_m^I \, , \quad 
\mu_{n m}^\pm = \lambda_m^R \lambda_n^I \pm \lambda_n^R \lambda_m^I \, , \\
\kappa_{n m}^\pm &= \left|\lambda_n \right|^2  \left|\lambda_m\right|^2  \pm q_n^2 q_m^2 \, , \quad 
\xi_{n m}^\pm =  q_n^2 \left|\lambda_m\right|^2 \pm q_m^2\left|\lambda_n \right|^2 \, , \\
\chi_n &= q_n^2 + \left(\lambda_n^I\right)^2 - \left(\lambda_n^R\right)^2 \, , \quad
\rho_n = 2 \lambda_n^R \lambda_n^I \, .
\end{split}\end{equation}

We can write the phases for the apparent horizon as
\begin{equation}
\delta^\pm_{\mathcal{H}_n \mathcal{H}^\prime_m} \equiv \frac{\alpha_{\mathcal{H}_n} \pm \alpha_{\mathcal{H}^\prime_m}}{2} + \Delta^{\pm}_{\mathcal{H}_n \mathcal{H}^\prime_m} \, ,
\end{equation}
where the $\alpha$'s are the phases of the gauge invariants at the horizon
\begin{equation}
H_n(r_h) = \mathcal{H}_n e^{i \alpha_{\mathcal{H}_n} / 2} \, ,
\end{equation}
and the $\Delta$'s are as follows
\begin{equation}\begin{split}
\text{tensor: }& \tan{\left(\Delta_{\mathcal{X}_n \mathcal{X}_m}^{\pm}\right)} = \frac{\mu_{n m}^\pm}{\nu_{n m}^\mp} \, , \\
\text{shear: }& \tan{\left(\Delta_{\mathcal{Y}_n \mathcal{Y}_m}^{\pm}\right)} = 0  \, , \\
& \tan{\left(\Delta_{\mathcal{U}_n \mathcal{U}_m}^{\pm}\right)} = \frac{\mu_{n m}^\pm}{\nu_{n m}^\mp}   \, , \\
\text{sound: }& \tan{\left(\Delta_{\mathcal{V}_n \mathcal{V}_m}^{\pm}\right)} = 0 \, , \\
&\tan{\left(\Delta_{\mathcal{W}_n \mathcal{W}_m}^{\pm}\right)} = \frac{\mu_{n m}^\pm}{\nu_{n m}^\mp} \, , \\
&\tan{\left(\Delta_{\mathcal{Z}_n \mathcal{V}_m}^{\pm}\right)} = \frac{2 \lambda_n^{I} \lambda_n^{R}}{\chi_n}\, , \\
&\tan{\left(\Delta_{\mathcal{Z}_n \mathcal{W}_m}^{\pm}\right)} =  \frac{\left| \lambda_n \right|^2 \mu_{n m}^- + q_n^2 \mu_{n m}^+}{\left| \lambda_n \right|^2 \nu_{n m}^+ - q_n^2 \nu_{n m}^+} \, , \\
&\tan{\left(\Delta_{\mathcal{Z}_n \mathcal{Z}_m}^{\pm}\right)} =  \, \\
&\!\!\!\!\!\!\!\!\!\!\!\!\!\!\!\!\!\! \frac{\mathcal{R} \left( \mu_{n m}^\pm \kappa_{n m}^-  +  \mu_{n m}^\mp \xi_{n m}^-\right) \!-\! 
\mathcal{Q} q_n q_m \left( \rho_n \chi_m \pm \rho_m \chi_n\right)}
{\mathcal{R} \left( -\nu_{n m}^\mp \kappa_{n m}^+  +  \nu_{n m}^\pm \xi_{n m}^+ \right) \!-\! 
\mathcal{Q} q_n q_m \left( \chi_n \chi_m \mp \rho_n \rho_m \right)} \, .
\end{split}\end{equation}
Note that in the vacuum limit $\mathcal{Q} = 0$ and the last expression reduces to 
\begin{equation}
\text{vacuum: } \tan{\left(\Delta_{\mathcal{Z}_n \mathcal{Z}_m}^{\pm}\right)} =
\frac{\mu_{n m}^\pm \kappa_{n m}^-  +  \mu_{n m}^\mp \xi_{n m}^-}{-\nu_{n m}^\mp \kappa_{n m}^+  +  \nu_{n m}^\pm \xi_{n m}^+} \, .
\end{equation}

Finally the amplitudes usually factorize as $c_{\mathcal{H}_n \mathcal{H}^\prime_m} = c_{\mathcal{H}_n} c_{\mathcal{H}^\prime_m}$ with
\begin{equation}\begin{split}
c_{\mathcal{X}_n} &= \left| \lambda_n \right|  \pi T  \Sigma_h^2 \mathcal{X}_n \, , \\
c_{\mathcal{Y}_n} &= \mathcal{Y}_n \, , \\
c_{\mathcal{U}_n} &= \left| \lambda_n \right|  \pi T  \Sigma_h \sqrt{Z_h}  \mathcal{U}_n \, , \\
c_{\mathcal{V}_n} &= \frac{1}{2} \Sigma_h \sqrt{Z_h}  \mathcal{V}_n \, , \\
c_{\mathcal{W}_n} &= \frac{1}{2} \left| \lambda_n \right|  \pi T \Sigma_h^2 \mathcal{W}_n \, , \\
c_{\mathcal{Z}_n} &=  \frac{c_n^{\mathcal{Z},\text{vac.}}}{\sqrt{\mathcal{R}} \left| \lambda_n \right|} \, 
\end{split}\end{equation}
except in the following three cases
\begin{equation}\begin{split}
c_{\mathcal{Z}_n \mathcal{V}_m} &=  \frac{\left|q_n\right|}{2 \sqrt{2(d-1)} \pi T \left| q_n^2 - \lambda_n^2 \right|} a_h^{(t)}{}^\prime \sqrt{\frac{\Sigma_h Z_h}{\pi T \Sigma_h^\prime}} c_{\mathcal{V}_n}\, , \\
c_{\mathcal{Z}_n \mathcal{W}_m} &= \frac{\left|\lambda_n\right| \Sigma_h \left( - 2 V_h^\prime + Z_h^\prime \left(a_h^{(t)}{}^\prime\right)^2\right)}{16 (d-1) (\pi T)^2\left| q_n^2 - \lambda_n^2 \right| \Sigma_h^\prime} c_{\mathcal{W}_n} \, , \\
c_{\mathcal{Z}_n \mathcal{Z}_m} &= c_{\mathcal{Z}_n} c_{\mathcal{Z}_m} \times \\
&\sqrt{\mathcal{R}^2 \left| \lambda_n \right|^2 \left| \lambda_m \right|^2 + \mathcal{Q} q_n q_m \left( \mathcal{Q} q_n q_m + 2 \mathcal{R} \nu_{n m}^- \right)}  \, .
\end{split}\end{equation}

%%%%%%%%%%%%%%%%%%%%%%%%%%%

%%%%%%%%%%%%%%%%%%%%%%%%%%%
\section{Positivity}\label{sec:Positivity}
%%%%%%%%%%%%%%%%%%%%%%%%%%%

In this section we show explicitly that the divergence of the entropy current for the apparent horizon given by the lowest two tensor perturbations will be positive definite.
We have checked numerically that this statement holds in every other channel as well as for the event horizon.
 
To begin we rewrite the tensor mode contribution of Eq. (\ref{eq:div_S}) in the form of Eq. (\ref{eq:div_S_positive}). For clarity we introduce the following notation,
\begin{align}
\Omega_n & = k_n x - \omega_n^{R} \, t + \alpha_n/2 \, , \\
\Theta_{n} & = \lambda_{n}^{R} \cos (\Omega_n )-\lambda_{n}^{I} \sin (\Omega_n) \, ,  \nonumber\\
\Xi_{n} & = \lambda_{n}^{I} \cos (\Omega_n ) +\lambda_{n}^{R} \sin (\Omega_n) \, . \nonumber \\
g_n & = \frac{c_n}{\sqrt{8 r_h \pi T \left| \lambda_n\right|^2(1+q_n^2)} } \, . \nonumber
\end{align}

The contribution of a pair of modes $\omega_n$, $\omega_m$ is given by
\begin{equation}\label{eq:combi}
\partial_{\mu} S^{\mu} |_{n m} =\mathcal{A}_{n n} + \mathcal{A}_{m m} + \mathcal{A}_{n m} + \mathcal{A}_{m n} \, ,
\end{equation}
using the $\mathcal{A}$ from Eq. (\ref{eq:divSmatter}).
Using the above notation we can recast this contribution in the following form,
\begin{align} \label{eq:div_S_positive}
\partial_{\mu} S^{\mu} |_{n m} = & + g_{n}^2 e^{2 \omega_{n}^{I}\, t}\left( q_{n}^2 \Theta_{n}^2+
   \left(2+q_n^2\right) \Xi_{n}^2 \right) \, \\
   & + g_{m}^2 e^{2 \omega_{m}^{I}\, t}\left( q_{m}^2 \Theta_{m}^2+
   \left(2+q_m^2\right) \Xi_{m}^2 \right) \, \nonumber \\
   & + 2 g_n  g_m \, e^{(\omega_{n}^{I}+ \omega_{m}^{I}) \, t} \bigg[f_1 \,\left(  q_n q_m \Theta_n \Theta_m \right)  \nonumber \\
   & + f_2 \, \left(\sqrt{2+q_n} \sqrt{2+q_m} \, \Xi_n \Xi_m \right) \bigg] \, , \nonumber
\end{align}
with functions $f_1$ and $f_2$ defined as,
\begin{align} \label{eq:f_function}
f_1 & = \frac{\sqrt{1+q_{n}^2} \sqrt{1+q_{m}^{2}}}{\left(1+q_{+}^{2}\right) \left(1+q_{-}^{2}\right)} \, , \\
f_2 & = \frac{
   \left(q_{+}^2+q_{-}^2+2\right) \sqrt{\frac{\left(1+q_{n}^2\right) \left(1+q_{m}^2\right)}{\left(2+q_{n}^2\right) \left(2+q_{m}^{2}\right)}}}{\left(1+q_{-}^2\right) \left(1+q_{+}^2\right)} \, .
\end{align}

We can note that $f_1$ and $f_2$ are bounded as $0 \leq f_1 \leq f_2 \leq 1$. The first two terms of Eq. (\ref{eq:div_S_positive}) are positive definite meaning that $\partial_{\mu} S^{\mu}$ will attain its minimum value when the last two terms have the largest negative contribution. Setting $f_1 = f_2 = 1$ we find that Eq. (\ref{eq:div_S_positive}) can be written as the sum of two squares, showing that we must have $\partial_{\mu} S^{\mu} \geq 0$ and completing the proof, 
\begin{align}
\partial_{\mu} S^{\mu} |_{n m} \geq & \, \left(g_{n} e^{ \omega_{n}^{I} \, t} q_{n} \, \Theta_{n} + g_{m} e^{ \omega_{m}^{I} \, t} q_{m} \, \Theta_{m} \right)^2 + \, \\
\big(g_{n} e^{ \omega_{n}^{I} \, t} &\sqrt{2+q_{n}^2} \, \Xi_{n} + g_{m} e^{ \omega_{m}^{I} \, t} \sqrt{2+q_{m}^2} \, \Xi_{m} \big)^2 \, 
\geq 0  \nonumber \, .
\end{align}
The leading contribution to Eq. (13) is given when $n=0$ and $m=1$.

%%%%%%%%%%%%%%%%%%%%%%%%%%%
\end{appendix}

\end{document}